\documentclass[amsmath, amssymb, showkeys, preprint, superscriptaddress]{revtex4-1}
\usepackage[pdftex]{graphicx}
\usepackage{dcolumn}
\usepackage{bm}
\usepackage{framed}

\begin{document}

\begin{framed}
This work has been submitted to the IEEE for possible publication.
Copyright may be transferred without notice, after which this version may no longer be accessible.
\end{framed}

\title[lockfet]{The Miniband Alignment Field-Effect Transistor:\\a superlattice-based steep-slope nanowire FET}

\author{Maarten Thewissen}
 \email{maarten.thewissen@imec.be}
 \affiliation{imec, Kapeldreef 75, B-3001 Heverlee, Belgium}
 \affiliation{Department of Physics, University of Antwerp, Groenenborgerlaan 171, B-2020 Antwerpen, Belgium}
\author{Bart Sor\'{e}e}
 \email{bart.soree@imec.be}
 \affiliation{imec, Kapeldreef 75, B-3001 Heverlee, Belgium}
 \affiliation{Department of Physics, University of Antwerp, Groenenborgerlaan 171, B-2020 Antwerpen, Belgium}
 \affiliation{Department of Electrical Engineering, KU Leuven, Kasteelpark Arenberg 10, B-3001 Heverlee, Belgium}
\author{Wim Magnus}
 \email{wim.magnus@imec.be}
 \affiliation{imec, Kapeldreef 75, B-3001 Heverlee, Belgium}
 \affiliation{Department of Physics, University of Antwerp, Groenenborgerlaan 171, B-2020 Antwerpen, Belgium}

\date{\today}

\begin{abstract}
This work investigates energy filtering in nanowires, where pass and stopbands are obtained by including superlattices in the wire.
When a pair of such superlattices is placed in series, each being controlled by a gate, it can act as a transistor where the (mis-)alignment of its minibands turns the device on (off).
It is shown that, in the ballistic current-regime, the transition between the on and off state occurs in a narrow gate-bias range, giving rise to sub-60 mV per decade switching behavior.
\end{abstract}


\maketitle

After five decades of Moore's law the bottleneck hampering further downscaling of chip sizes is no longer process technology, but rather power consumption.
To reduce dynamic power the chip's supply voltage must be scaled down as well, thereby causing an exponential increase in leakage current if the field-effect transistor's (FET) threshold voltage is to be scaled along.
The resulting trade-off between power consumption and performance effectively follows from the impossibility to turn off transistors abruptly, as represented by a subthreshold slope of at least 60 mV per decade of current. As such, improving this important figure of merit is urgently required for future devices.

A possible solution starts from the observation that most of the subthreshold current comes from high-energy carriers that remain present according to Fermi-Dirac statistics.
Preventing these carriers from being injected in the device should then suppress this current, an idea known as energy filtering.
A familiar representative of energy-filtering devices is the Tunnel FET (TFET).
Here, electrons with energies in the tail of the Fermi-Dirac distribution are filtered out by the band gap of the source material. 
A disadvantage of this approach is that band-to-band tunneling is required to turn on the device, which severely limits the on-current~\cite{seabaugh2010low}.

Another concept exploiting quantum mechanics for energy filtering is based on superlattices~\cite{gnani_performance_2011}.
Resonance in a superlattice creates minibands and miniband gaps, the latter filtering out the unwanted high-energy electrons.
Although comparatively less examined, these devices are in principle more promising than the TFET~\cite{gnani2016steep}.

\begin{figure}[!ht]
\centering
\includegraphics{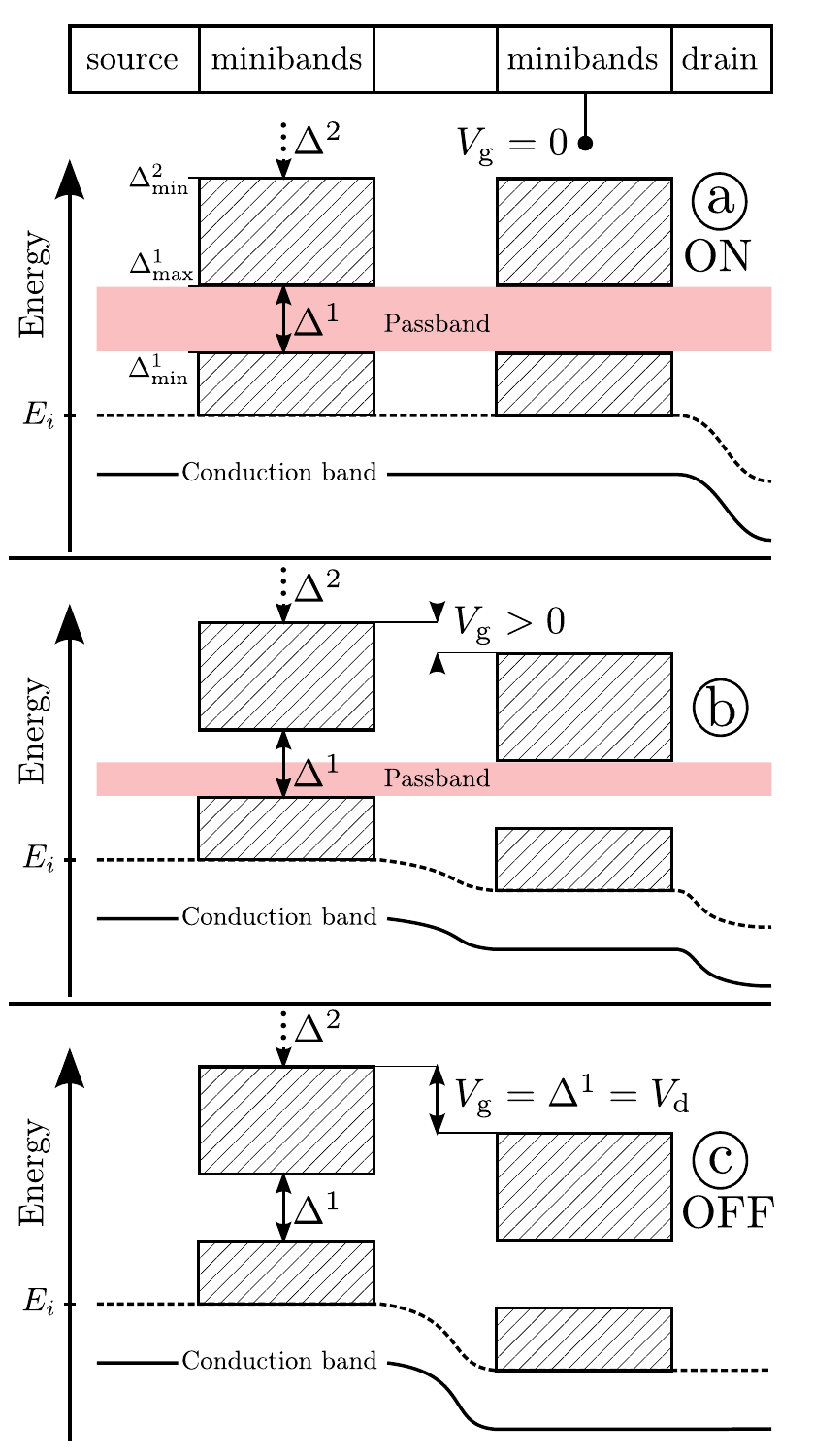}
\caption{Working principle of the MiAFET explained schematically, depicting the two lowest minibands (with bandwidths $\Delta^1$ and $\Delta^2$) for a subband $i$ (with confinement energy $E_i$). 
\mbox{(Mis-)alignment} of the minibands turns the device on (off).}
\label{fig:schematic}
\end{figure}

In this work, we introduce a novel device concept based on this idea: the Miniband Alignment FET (MiAFET).
Two superlattices are placed in series, each giving rise to a local miniband spectrum that acts as an energy filter.
Fig.~\ref{fig:schematic} shows schematically how efficient switching can be realized in this configuration.
Where the minibands of both superlattice regions align, passbands are formed and current can pass.
Complete alignment, corresponding to the maximal current case, is pictured in Fig.~\ref{fig:schematic}a.
When the potential in the second superlattice increases, the passbands become more narrow, resulting in a decrease of the current (Fig.~\ref{fig:schematic}b).
As soon as misalignment of the lowest minibands is complete, the lowest passband ceases to exist and the current drops sharply (Fig.~\ref{fig:schematic}c).
Because the tail of the Fermi-Dirac distribution is blocked ``from above'', states carrying the most current are filtered abruptly, and a turn-off slope much steeper than 60 mV per decade can be realized.
The remaining off-current arises from the second minibands $\Delta^2$ in both superlattice regions, that are still partly aligned.
Their contribution, however, can be minimized by ensuring that $\Delta^2_{\mathrm{min}}$ occurs at a sufficiently high energy, where the occupation probability is negligible.
Another restriction on the miniband spectrum is that the second miniband gap, $\Delta^2_{\mathrm{min}} - \Delta^1_{\mathrm{max}}$, must be wider than the lowest miniband $\Delta^1$ to ensure that complete misalignment can be realized.
From this simple picture, it also follows that the minimal required drain bias equals $\Delta^1$, the gate bias varying from 0 V to this drain bias.

\begin{figure}[!ht]
\centering
\includegraphics{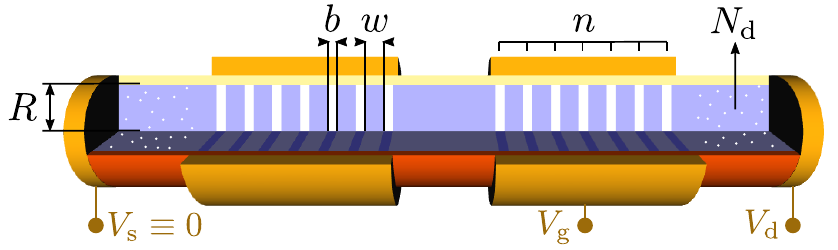}
\caption{Proposed device configuration, most important parameters being the wire radius ($R$), the barrier and well widths ($b$ and $w$), the number of barriers ($n$) and the doping in source and drain ($N_\mathrm{d}$).}
\label{fig:lockfet_3d}
\end{figure}

As an implementation of this idea, consider the device configuration in Fig.~\ref{fig:lockfet_3d}: a pair of superlattices is incorporated in a nanowire, both surrounded by a gate.
Only the second gate is contacted however, the first one only serving the purpose of electrostatic doping as an alternative to impurity doping~\cite{maiorano_design_2014}.
Consequently, the superlattices and the region in between are undoped.
For the sake of illustration we choose the material pair in this paper to be lattice matched GaAs/AlGaAs, which has a conduction band offset of 0.5 eV and an effective mass of $0.063\,\mathrm{m}_\mathrm{e}$.
The barrier and well width $b$ and $w$ are chosen to be 1.5 nm and 4.0 nm, respectively, leading to $\Delta^1 \approx 0.11$ eV.
In our examples, $V_\mathrm{d}$ is therefore set to 0.15 V, i.e. slightly larger.
In what follows, we will investigate the variation of the number of barriers $n$, the radius $R$ and the lead doping $N_\mathrm{d}$.
Unless otherwise stated, $n = 8$, $R = 5\  \mathrm{nm}$ and $N_\mathrm{d}$ is chosen such that the Fermi level in the source coincides with $\Delta^1_{\mathrm{min}}$ of the lowest subband.
All simulations assume an $\mathrm{Al}_2\mathrm{O}_3$ oxide of 1 nm (dielectric constant $\epsilon \approx 10$, electron effective mass $\mathrm{m}^*_{\mathrm{e}} \approx 0.4\, \mathrm{m}_{\mathrm{e}}$~\cite{perevalov_electronic_2007}, energy barrier w.r.t GaAs $\approx 3.0$ eV~\cite{nguyen_band_2008})

In our simulations, only the conduction band is taken into account in view of the high doping concentration in source and drain.
The energy eigenfunctions of the device are obtained by solving the time-independent Schr\"{o}dinger equation with quantum transmitting boundary conditions~\cite{lent_quantum_1990}.
Deriving the charge density from these states, Poisson's equation yields a new potential and the process is repeated until a self-consistent potential is obtained.
All partial differential equations (2D, because of the axial symmetry) are solved using the finite-element software FEniCS~\cite{logg_automated_2012}. 
As to the steady-state occupation of the states, we adopt ballistic transport assuming that a state is occupied according to the Fermi-Dirac distribution of the lead it originates from. 
This leads to the following expression for the current:

\begin{equation}
I = \frac{\mathrm{e}}{\pi \hbar} \sum_{i}^\infty \int_{E_{i}}^\infty T_{i}(\varepsilon) [f(\varepsilon, \mu_\mathrm{s}) - f(\varepsilon, \mu_\mathrm{d})] \, \mathrm{d} \varepsilon
\label{eq:current}
\end{equation}
where $T(\varepsilon)$ is the transmission coefficient written as function of electron energy, $f$ is the Fermi-Dirac distribution and $\mu_\mathrm{s}$ and $\mu_\mathrm{d}$ respectively denote the Fermi-level in source and drain.
The summation runs over all subbands that are present due to radial confinement.

\begin{figure}[!ht]
\centering
\includegraphics{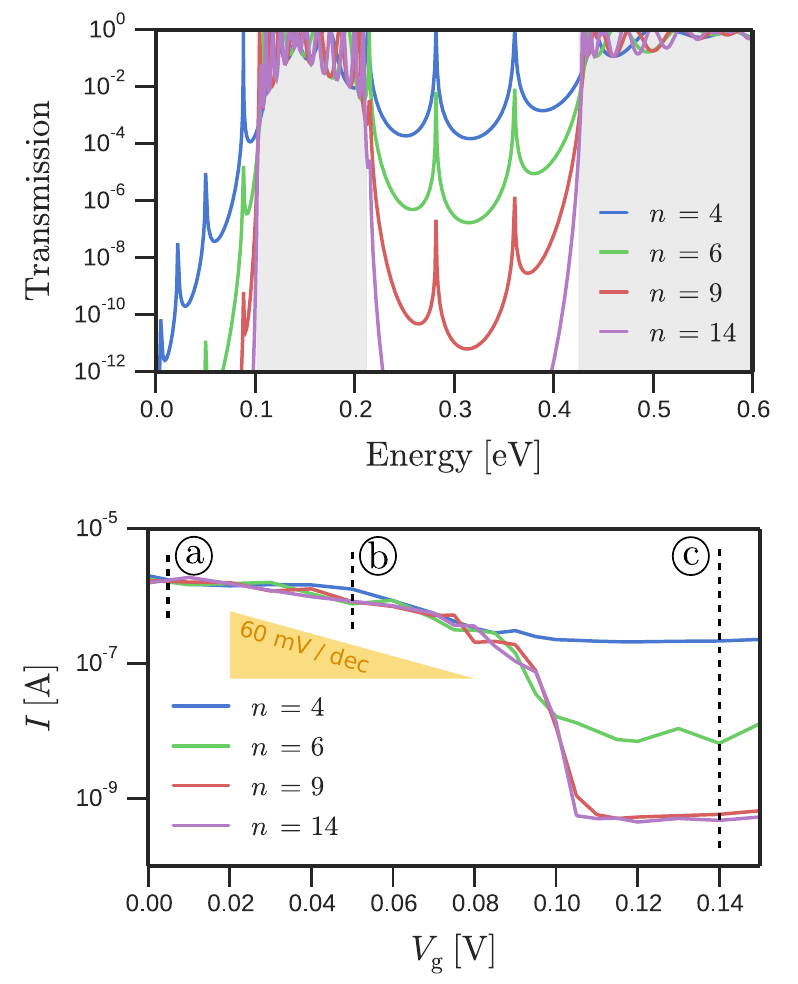}
\caption{
(top) Transmission spectrum of the unbiased device for a varying amount of superlattice barriers $n$. The minibands of the Kronig-Penney model for the same values $b$ and $w$ are shaded in grey. 
(bottom) Corresponding IV-characteristics. For three different cases \textcircled{a}, \textcircled{b} and \textcircled{c}, the idealized and simulated miniband structures are plotted in Fig.~\ref{fig:schematic} and Fig.~\ref{fig:ldos}, respectively.
}
\label{fig:varying_nrbarriers}
\end{figure}

Fig.~\ref{fig:varying_nrbarriers} shows how the transmission spectrum and the current-voltage characteristics are affected when $n$, the number of barriers, is varied.
For increasing $n$, the transmission minibands converge to those of an infinite superlattice, as appearing in the Kronig-Penney model~\cite{kronig1931quantum, liboff1980introductory}:
single transmission peaks that correspond to resonance between---rather than within---the superlattices disappear and transmission in the miniband gaps drops to insignificant low values.
For superlattices containing only 3 or 4 barriers, on the contrary, transmission in the miniband gap is too significant for a steep subthreshold-slope to emerge.
Inclusion of more barriers was found to result in more clearly defined minibands and a sharper current drop, exceeding one decade over 10 mV, as predicted by the simulations.
From Fig.~\ref{fig:varying_nrbarriers} it can be seen that, beyond 9 barriers however, further improvement of the subthreshold slope is marginal.
It is worth noticing that increasing the applied gate bias beyond $\Delta^1 \approx 0.11$ eV has little effect, as the minibands cannot be more misaligned as they already are.

\begin{figure}[!ht]
\centering
\includegraphics{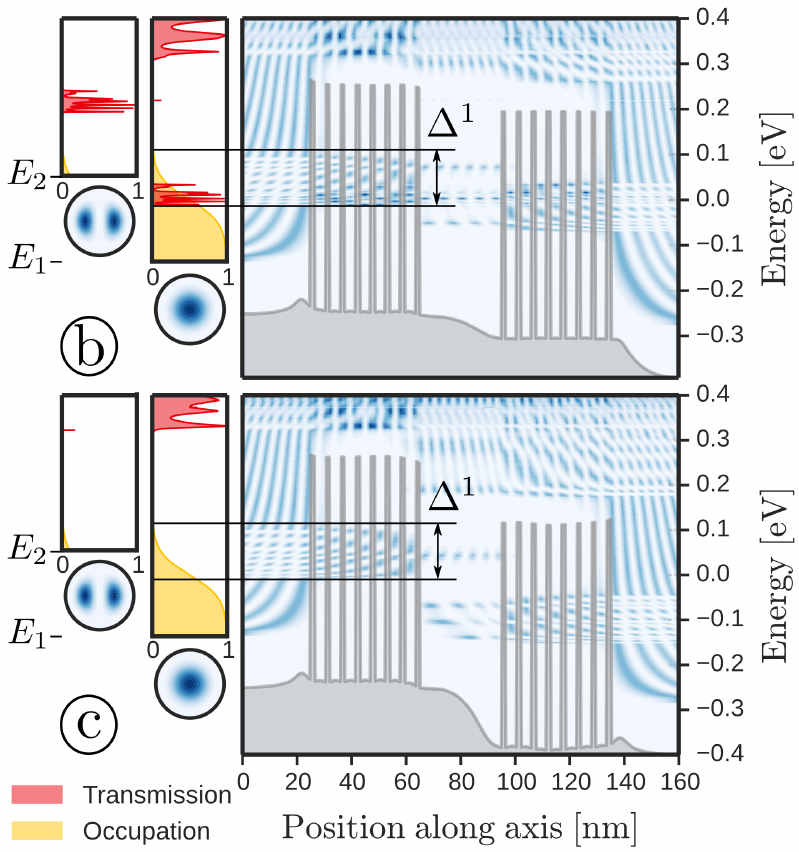}
\caption{
Transmission spectrum and occupation probability for the lowest two subbands in the source, and the subband's electron density in a cross-section of the lead. 
For the lowest one, the local density of states $\scriptstyle{\left| \Psi \left( x; \scriptscriptstyle{E} \right) \right|^2}$ is also shown together with the conduction band minimum.
}
\label{fig:ldos}
\end{figure}

For specific gate biases of 0.05 V and 0.14 V, Fig.~\ref{fig:ldos} shows the simulated miniband structure.
The transmission spectrum and occupation probability are plotted, the ballistic current being determined by the overlap of both according to Eq.~\ref{eq:current}.
From Fig.~\ref{fig:ldos}b (top) it is clear that the lowest subband dominates the on-current, whereas in Fig.~\ref{fig:ldos}c (bottom) the remaining current originates from the second miniband.
As explained earlier, the low occupation probability of these states ensures that the corresponding off-current is small.
Whether resonance occurs or not depends on the electron wave vector perpendicular to the superlattice planes.
Consequently, the transmission spectra for all subbands are similar, differing only by an offset due to a different confinement energy.
The right side of Fig.~\ref{fig:ldos} should be compared with the idealized scheme in Fig.~\ref{fig:schematic}:
inside the superlattice regions, one can clearly identify the local minibands and miniband gaps, an overlap of minibands resulting in a passband with nonzero transmission.

\begin{figure}[!ht]
\centering
\includegraphics{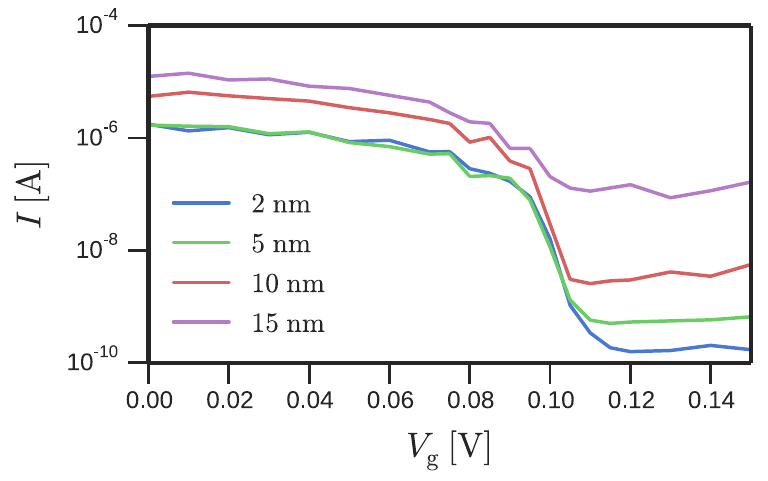}
\caption{IV-characteristics for a varying wire radius R.}
\label{fig:varying_diameter}
\end{figure}

If $R$ is small enough, the potential varies only notably in the transport direction, ensuring that the transmission spectra for all subbands are similar.
For increasing wire radii $R$, weaker energy quantization causes more subbands coming into play, and the current is seen to increase (Fig.~\ref{fig:varying_diameter}).
Moreover, when $R$ is taken to be larger, the influence of the gate is most pronounced near the surface, and the potential variation in the radial direction is no longer negligible.
As a result, lower subbands with a higher electron density near the symmetry axis of the wire will experience a different conduction band offset in the superlattice.
Transmission spectra will vary between subbands and misalignment no longer occurs at the same gate bias, if at all.
Hence, it becomes more difficult to turn off the device, although sub-60 mV per decade slopes are still possible for wire radii up to 10 nm, as shown in Fig.~\ref{fig:varying_diameter}.

\begin{figure}[!ht]
\centering
\includegraphics{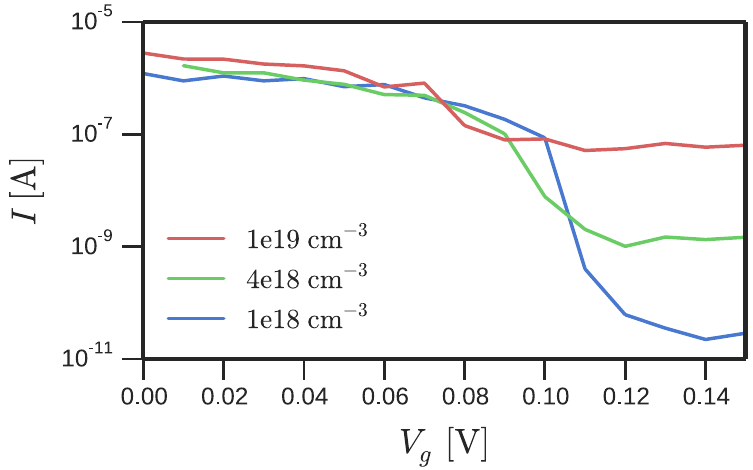}
\caption{IV-characteristics for varying lead doping $N_\mathrm{d}$.}
\label{fig:varying_leaddoping}
\end{figure}

Another important parameter is the doping $N_\mathrm{d}$ of the source and drain leads.
The lead doping together with the wire radius determines the position of the Fermi-level with respect to the minibands.
Fig.~\ref{fig:varying_leaddoping} shows that, as expected, the current increases with increasing doping levels.
But the current contribution of the second miniband $\Delta^2$ also becomes larger and, as mentioned earlier, these minibands do not misalign and set a lower limit for the off-current.
This leads to a trade-off between high doping levels to achieve a substantial on-current, and low occupation of the higher minibands to prevent the device from turning off properly.
In practice, good results are obtained when the Fermi-level of the source lies within the lowest miniband.
The optimal doping level depends therefore on both the radius $R$ and barrier width $b$ and well width $w$.
Varying the doping within the superlattices or in the region between them was found to have little influence, because---at least for the diameters considered here---the electrostatic potential is dominated by the gate around it.
The situation would be different when we remove the gate around the first superlattice.
In that case the selfconsistent potential for which the device still operates would strongly depend on the doping level in this first superlattice.

In conclusion, as the limits of process technology are pushed onwards, newly created devices enter the realm of quantum mechanics. 
While being a nuisance in conventional MOSFETs, this also opens possibilities for new device concepts that rely on e.g. resonant tunneling to operate. 
It was shown that a pair of built-in gate-controlled superlattices placed in series can lead to alternative transistor structures, achieving acceptable on-currents and a subthreshold slope less than 10 mV/decade.
No effort was made yet to optimize w.r.t. the wire radius, barrier and well widths or lead doping.
The device is in principle less susceptible to surface scattering than TFETs, because the electron wave functions do not pile up near the oxide interface, but are spread out over the device.
Also, no impurity dopants are present in the active device regions that can destroy coherence of the wave functions.

\nocite{*}
\bibliography{article}

\end{document}